# What is balance? A vital mechano-regulation paradigm

Author: Nicholas M. Wilkinson

Affiliation: Independent researcher

ORCID: https://orcid.org/0000-0001-8539-910X

**Abstract**
Within minutes of birth a newborn gnu or giraffe works to stand and walk, asserting postural balance and organised animate behaviour in an apparently goal-directed manner. In contrast, robots learning to stand and walk from scratch begin with random flailing, the behaviour cohering over time as the robot internalises some reward/value signal. How does the newborn gnu 'innately know' what goal to aim for, and decide to work towards it? How could similar goal-directed balance learning be implemented in robots? Currently, animate balance inherits its axiomatic definition from the Newtonian formulation for inanimate balance; static mechanical equilibrium. This is arguably inappropriate for animate balance, because animals need to move and are never in static mechanical equilibrium, giving rise to the 'posture-movement paradox'. The present Perspective proposes a more fluid, dynamical axiomatic task definition and goal which (a) isolates resisting gravity, (b) admits and enables movement, and (c) subsumes static mechanical equilibrium as a special case. This novel definition is founded upon inevitable biophysical requirements and observable developmental process. The article explains how animals apprehend and embed this goal through prenatal development suspended in equidense amniotic fluid, and then are challenged to self-maintain it by the perinatal transition. The account entails a paradigmatic shift in putative physiological organisation and associated conceptual framework for balance; from a subsidiary sensorimotor control task to a vital mechano-regulation task, organisationally akin to thermo-regulation. This vital mechano-regulation model of balance has practical implications and implies a range of predictions.

## Introduction

Resisting gravity is a fundamental and constant requirement for life on Earth (Anken & Rahmann, 2002; Vinogradova et al., 2021). 'Antigravity' movements are a standard diagnostic of newborn human health (Miyagishima et al., 2016; Piper et al., 1992). Many newborn ungulates rapidly and deliberately work to stand and balance minutes after birth (Vanden Hole et al., 2017). In contrast, robots learning to balance from scratch start with random flailing, and gradually improve in response to a reward signal or other feedback (Liang et al., 2018; Radosavovic et al., 2024; Rieffel & Mouret, 2018). How is this task defined in the newborn animal (Körding, 2007; Lyon & Kuchling, 2021), such that they seem to know what to do from the start, whilst robots currently do not? Is this goal-directed behaviour (Di Paolo, 2010; Heylighen, 2023)? If so, what is the goal, how is it established and embedded developmentally, and could the same goal be instantiated in robots? How could a developmental system become organised around an intrinsically unstable, non-equilibrium goal (Heylighen, 2023)? The present contribution addresses these questions, proposing a novel axiomatic task definition for animate balancing. Apprehending the argument requires confluence of multiple novel concepts. As a result, the reader may find that some aspects become clearer in hindsight, once other aspects are communicated. We have tried to minimise this barrier to communication via the order of presentation, but have not been able to eradicate it.

For inanimate matter, balance is defined in Newtonian mechanics as mechanical equilibrium; the state where the load actions (forces or moments) acting upon a massive body sum to zero (Pollock et al., 2000), so it stays still in the local frame of reference (Figure 1a). Static equilibrium also provides an axiomatic theoretical foundation for animate balance; 'balance' and 'equilibrium' are typically used interchangeably (Pollock et al., 2000). However, living animals are never in this state due to their animacy, non-rigidity, and dynamic internal states (Kelso, 2009). Nevertheless, animals are usually balanced (Figure 1b), even adapting to radical perturbations of morphology (Figure 1c). As a result, an intuitive notion of 'not-falling', usually termed 'stability', stands in for a formal definition (Le Mouel & Brette, 2017). The equation of balance with stasis gives rise to the 'posture-movement paradox' (Ivanenko & Gurfinkel, 2018; Latash, 2025; Massion et al., 2004; Von Holst & Mittelstaedt, 1950); why does force for balance not impede force for movement? Extant work-arounds include 'efference copy' (Grüsser, 1994), and 'equilibrium point control' (Feldman & Levin, 2009). The former involves message-passing to inform relevant controllers about intended movements, such that they can be accommodated. The latter resets the default posture at which stasis should be achieved. It has also been proposed that the goal/purpose of balance is not static equilibrium but mobility (Le Mouel & Brette, 2017). This approach shares aspects with the current proposals, but skilled use of body weight and pendular action for movement is not present at birth and develops over time (Ivanenko et al., 2007; Le Mouel & Brette, 2017), arguably implying a prior purpose/goal; this prior 'innate' goal is our target here.

The present contribution builds on (Wilkinson, 2008), proposing that animate balance requires a more fluid task definition which isolates work resisting gravitational acceleration specifically, whilst admitting other movement. Specifically, balance (noun) is a particular field of mechanical force; i.e. that which precisely cancels gravitational acceleration for a given body. Balancing (verb) is the self-regulation of this force field, and as skill develops the utilisation of small perturbations to it for movement (Le Mouel & Brette, 2017). This force field is fluid and dynamical, being different for each morphology and configuration thereof, and so may appear impossible to define in a general manner. Nonetheless, there is a way to define it for arbitrary morphology and dynamical configuration, which is explicated in the argument here.

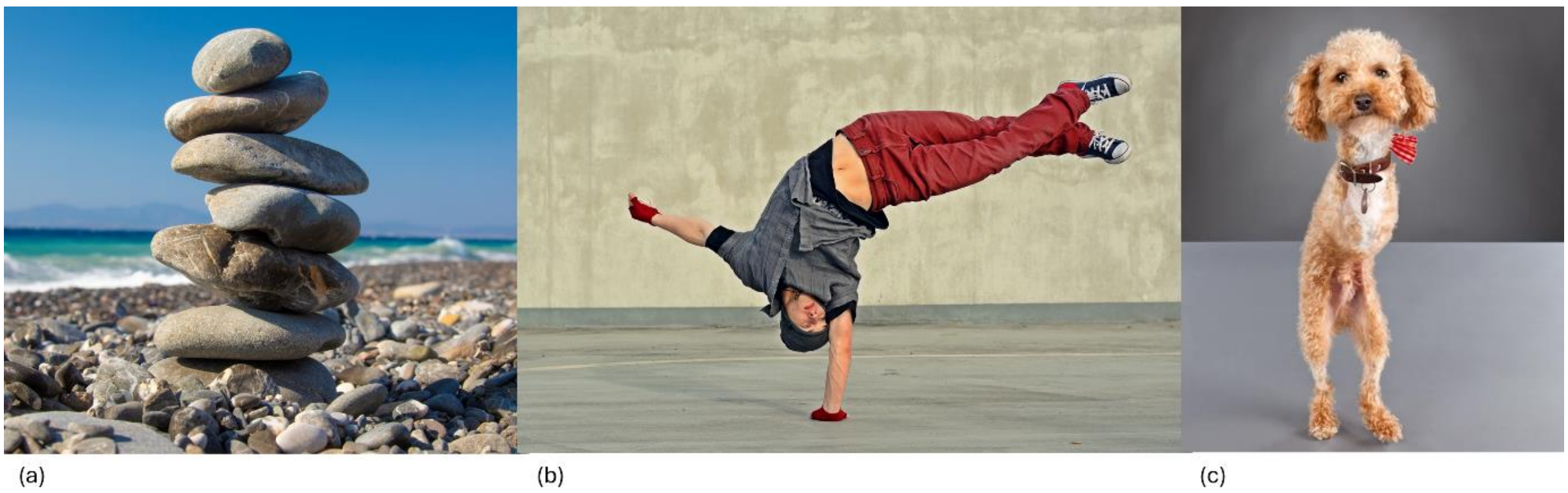

Figure 1. *(a) Inanimate balance is characterised by static stability. Image credit* Petr Kratochvil *Public Domain* ***CC0 1.0 Universal*** *(b) Animate balance is characterised by dynamical stability. Image credit* chphotography85*, Adobe Stock Images (Standard license) (c) Animal balance can adapt to perturbations of phenotype, so explanations should accommodate this adaptiveness and flexibility. Image credit* Carli Davidson *(personal permission)*

Currently, balancing is taken to be a subsidiary sensorimotor control task whose goal is ancillary support to volitional action (Ivanenko & Gurfinkel, 2018; Le Mouel & Brette, 2017; Massion et al., 2004) (counter-examples to this framing are invited). The present contribution questions this paradigmatic framing, suggesting that balance is instead organised as a vital task-based modality (organisationally akin to thermo-regulation) in the domain of mechano-regulation (Turvey & Fonseca, 2009). Specifically, whole-body balance actuates mechanical force to compensate the density gradient between body and medium, which is close to zero during early prenatal development but increases radically over the perinatal transition.

The reader is urged to familiarise with the definition of terms. Section 1 reviews the norms currently thought to drive effortful gravity resistance in animals and used to drive balance in robots. Section 2 briefly introduces homeostasis and vital goal-directed systems, and how these may relate to the development of balancing. Section 3 outlines the key concept of buoyancy and defines usage here in formal and conceptual detail. Section 4 describes the timeline and significance of amniotic buoyancy regulation in human pregnancy. Section 5 defines a high-level model of balance as vital mechano-regulation. Section 6 discusses evidential fit, limitations and scope of this model for understanding the development of balance and spatiotemporal animacy in animals and prospectively outlines how these ideas could be applied to robot balance.

**Definition of terms**

**Mechano-regulation:** self-regulation of mechanical force.
**Normative:** the domain of deeming things (actions, sensations, predictions) good versus bad. There are many related terms, including motivation, drive, value function, reward function, objective function, control policy.
**Norm**: A value or state against which things are deemed good or bad, including goals, set-points, reward delivery conditions, error signals and the like.
**Goals and set-points:** *normative values* defining what level some variable(s) *should* be at. Sometimes these terms are used interchangeably, but here they hold distinct meanings.
**Goal**: denotes an *objective world state* normative value (not a representation), whose normative feedback is provided *by physics.* Body temperature being ~37$^{o}$C is a goal.
**Set-point:** denotes a *control system* normative value, whose feedback is enacted *by the control system*. Set-points can be representational e.g. the temperature dial on a thermostat or a neural representation of a target temperature in an animal.
**Proxy:** Set-points can act as *proxies* for goals, if they are appropriately calibrated (John et al., 2024). For example a 20$^{o}$C *set-point* on a home heating thermostat is a *proxy* for the *goal* of a physically warm home.
**Control policy:** the normative organisation of control in an animal or robot which defines what it ‘should’ do, for example by calibrating its set-points to proxy a particular goal, or by designing it to maximise a given reward and defining the reward delivery conditions.

## 1. Task definition for balancing

*"The term balance (or equilibrium), as used in mechanics, is defined as the state of an object when the resultant load actions (forces or moments) acting upon it are zero (Newton's First Law)…The principles of Newtonian mechanics…are equally applicable to the balance of humans (or animals) as they are to inanimate objects."*
Pollock, Durward and Row (2000) "What is balance?" pp402

The current overarching definition of balance is static mechanical equilibrium; 'balance' and 'equilibrium' are typically used interchangeably. Multiple theoretical paradigms for sensorimotor control are in extant use, with varying perspectives on how balance fits in (Feldman & Levin, 2009; Fullerton, 2025; Gage et al., 2004; Gurfinkel et al., 1995; Ivanenko & Gurfinkel, 2018; Massion & Woollacott, 2004; McNamee & Wolpert, 2019; Morasso et al., 2019; Silva et al., 2010; Turvey & Fonseca, 2009). Many aspects of sensorimotor competence in animals result from muscle synergies and embodied intelligence, rather than neural control alone (Adolph & Hoch, 2019; Craighero, 2024; Kelso, 2009; Latash, 2010; Turvey & Fonseca, 2009, 2014; Zhao et al., 2024). The proposals here apply equally to the developmental calibration of such embodied control systems as to neural ones. It is impossible to do justice to this extensive literature in the present contribution, which aims only to characterise how existing approaches conceive the normative task definition and control policies (Körding, 2007) driving balancing effort. In this domain, there is a ballpark consensus. As Massion and Woollacott (Massion & Woollacott, 2004) explain in their introduction, the task is typically decomposed into three subtasks; (1) asserting postural uprightness (Barbieri et al., 2008; Barra et al., 2010; Dakin & Rosenberg, 2018; Massion & Woollacott, 2004; Schoenmaekers et al., 2025), (2) maintaining the centre of mass within the vertical projection of the base of support (henceforth 'CoM/BoS') (Pollock et al., 2000; Van Dieën et al., 2024), and (3) support for desired target postures for action and the interactions they afford (Ivanenko & Gurfinkel, 2018; Massion & Woollacott, 2004; Mullick et al., 2018; Pollock et al., 2000; Van Dieën et al., 2024).

Early research posited that animals are reflexively driven to assert uprightness by a genetically specified reference posture; upright stance (Magnus, 1926; Massion et al., 2004; Rademaker, 1947; Sherrington, 1915). Whilst subsequent results suggest that reflexive movement is insufficient to account for the flexibility of balancing behaviour (Massion & Woollacott, 2004), the concept of postural uprightness/verticality remains central to defining the motivation for the 'get-up' aspect of balancing (Barbieri et al., 2008; Dakin & Rosenberg, 2018; Mullick et al., 2018; Schoenmaekers et al., 2025). It is often used in robotics, where the human user typically defines what will count as 'upright' for the robot (Melnyk & Pitti, 2018; Morimoto & Doya, 2001; Yoshida et al., 2021).

However, a set 'upright' reference posture struggles to explain the adaptivity of balance (Figure 1c), and animals balance when upside down too (Figure 1b). Balancing can be dissociated from verticality by manipulating loading (Dietz et al., 1992; Riccio et al., 1992). In utero, foetuses do not prefer an 'upright' position, though they easily could (Cunningham et al., 2010). This suggests that the sensitivity of balance to verticality is gated by loading (Dietz et al., 1992). Maintaining CoM/BoS relations describes a physically necessary sub-task to avoid falling (Pollock et al., 2000). However, it does not fully specify whole body balancing. For example, supporting reaching is taken to be a separate task, sitting upon the stable foundation of CoM/BoS posture (Hardesty et al., 2023; Ivanenko et al., 2011; Khatib et al., 2022; Rochat, 1992). CoM/BoS stability also assumes the rigidity needed to distribute mechanical force from centre of mass to extremities, limiting the generality of the approach for non-rigid bodies. CoM/BoS relations and uprightness are complicated for an agent to calculate from sensory data, necessitating the pre-existence of complicated internal self-models (Ivanenko & Gurfinkel, 2018; Massion, 1998). In robotics these models can be implemented by human hand, but the phenotypic norms constraining the development of these putative models in animals are not known. Finally, the top-level operational goal of balance is achieving and supporting the postural targets of volitional action (Feldman, 2024; Ivanenko & Gurfinkel, 2018; Massion & Woollacott, 2004; Mullick et al., 2018; Pollock et al., 2000; Van Dieën et al., 2024). However, volition is amongst the least well understood concepts in psychology and neuroscience so this dependency is a key weakness, as has been noted elsewhere (Turvey & Fonseca, 2009).

An alternative approach within robotics is to contain task definition for balancing within the definition of other tasks such as locomotion, with the above sub-tasks subsumed in the rewarded task (we are not aware of a theory of biological balance based on this approach). For example, locomotion (including balance) can be learned through reinforcement learning by including distance-travelled in the reward function (Liang et al., 2018; Radosavovic et al., 2024; Rieffel & Mouret, 2018), though complex robot bodies can require extensive hand-tuning of the reward function (Ibarz et al., 2021; Radosavovic et al., 2024). In these models the early learning period is characterised by random flailing; very different to the directed efforts observable in a newborn ungulate working to stand and walk (Vanden Hole et al., 2017). This approach results in a model of balance which is task-specific and so not easily transferable to other tasks. Neither of the above frameworks quite explains what control policy drives the directed learning of balanced stand-and-walk behaviour in a newborn ungulate. As a result, this apparent 'drive' is often designated innate; "*There is consensus that posture and locomotion are examples of such basic, genetically determined behaviors.*" (Massion et al., 2004) pp.15. This labels the behaviour but does not explain it and a rigid genetic definition may be inconsistent with the adaptivity of posture and balance (Figure 1c). We propose instead that the balancing task is determined biophysically and apprehended *ontogenetically* by a particular developmental process (though naturally under genetic constraints).

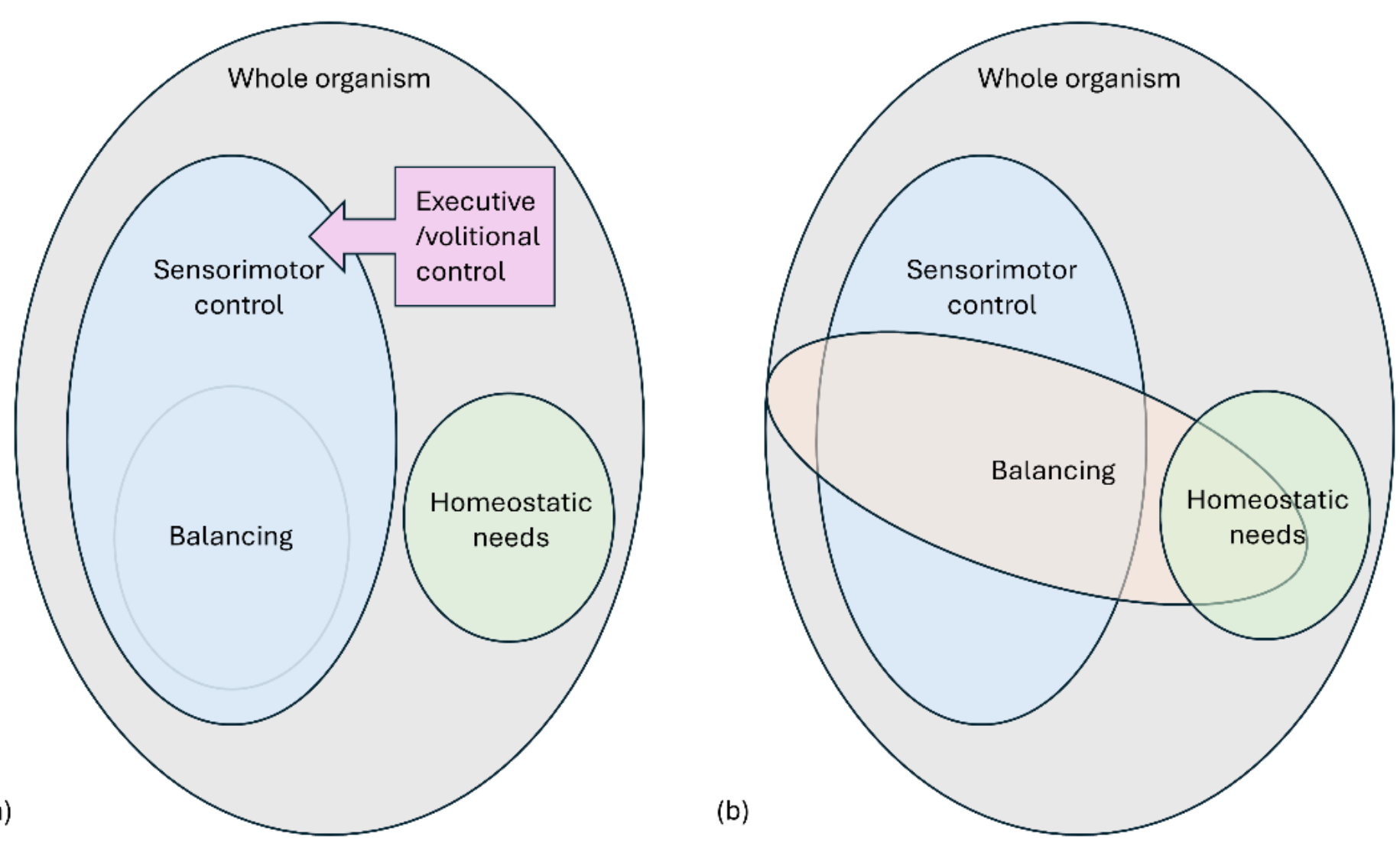

Figure 2 (a) *Balancing is currently conceived as a subsidiary sensorimotor control task. (b) We propose an alternative model of balance as a vital task-based modality of mechano-regulation, organisationally akin to thermo-regulation.*

Overall, balancing is currently taken to be a set of subsidiary sensorimotor control tasks providing ancillary support to the goals of volitional action. Balance is a consequence, *but not a cause*, of this subsidiary activity. Figure 2a depicts a crude boxological visualisation of this current conceptual framework (2a), against which to contrast the alternative framework proposed here. On this view, gravity resistance (i.e. balance) is a vital task-based modality driven by homeostatic need, organisationally akin to thermo-regulation i.e. organised around conserving a vital physical condition (2b). Balancing recruits sensorimotor components but is not limited to them. It is not driven by volitional action, but is necessary to all action, just as thermo-regulation is not driven by volitional action but is necessary to all action. The next section briefly outlines homeostasis as pertains to the origins and enaction of goal-directedness in the context of the current argument.

## 2. Homeostasis, 'co-homeostatic handover', and the origin of vital goals

The original concept of homeostasis as self-regulation of the internal milieu (Cannon, 1929) was reduced to reactive servo-control on set-points for essential variables in cybernetic formulations (Ashby, 2013; Ashby, 1960). Contemporary views are more refined and expansive (Bettinger & Friston, 2023; Ikegami & Suzuki, 2008; Keramati & Gutkin, 2014; Pezzulo et al., 2015; Tschantz et al., 2022; Yoshida, Daikoku, et al., 2024), as indeed are critiques (Barrett, 2020a, 2020b). Here we use 'homeostasis' in the specific sense of anchored in the need to maintain *vital* biophysical parameters within *set* viability bounds, but conceived expansively (Bechtel & Bich, 2024b) as a fundamentally dynamical process involving sensorimotor coordination (Ikegami & Suzuki, 2008; Soodak & Iberall, 1978), and inclusive of extended forms of homeostasis such as ultrastability (Ashby, 1960; Cariani, 2009), homeorhesis (Mamontov, 2007) and aspects of allostasis (Bettinger & Friston, 2023; McEwen & Wingfield, 2003; Sterling, 2012).

The canonical example of organism-level homeostasis is thermo-regulation (Benzinger, 1969; Cannon, 1929; Tan & Knight, 2018). The *goal* of a viable (non-equilibrium) body temperature is *proxied* (John et al., 2024) by many bodily *set-points* which flag deviation from (and threats to) the goal temperature range (see definition of terms). Actions are deemed 'good' or 'bad' by *normative* feedback on the status of the vital goal (Körding, 2007). Importantly, vital homeostatic goals are both *cause and consequence* of lower level self-regulatory activity (Di Paolo, 2010). For example, a drop in body temperature causes shivering, which in turn raises body temperature (Tan & Knight, 2018). This closed-loop macro-organisation helps to coordinate diverse subcomponents into a coherent behaviour, in the service of a shared goal (Bechtel & Bich, 2024b; Benzinger, 1969; Bettinger & Friston, 2023; Cannon, 1929; Ikegami & Suzuki, 2008; Keramati & Gutkin, 2014; Tan & Knight, 2018; Yoshida, Daikoku, et al., 2024).

Homeostasis theory addresses the persistence of a vital far-from-equilibrium goal state (Heylighen, 2023), but not its origin. Some bootstrapping is required to *establish* the goal state and *calibrate* bodily set-points to effectively proxy it (John et al., 2024). One potential developmental route we term 'co-homeostatic handover'. Mammalian homeostasis is achieved developmentally through 'co-homeostasis' within another supporting body (Ciaunica et al., 2021). For example, the umbilical cord supplies oxygenated blood and nutrients (Spurway et al., 2012), and the parental body insulates the foetus against temperature gradients (Underwood et al., 2005), giving the foetus time and resources to develop its own self-regulatory capacities. At birth, this insulation of the parental body is withdrawn, 'handing over' the task to the neonate and hence challenging it to self-regulate, though caregivers may continue to assist (Ciaunica, 2019; Fotopoulou & Tsakiris, 2017). Related notions of adaptive development in response to challenge to habituated norms have been explored in simulated agents (Di Paolo, 2010) and active matter (Shreesha & Levin, 2024). The present argument suggests that the drive to balance is embedded in a similar developmental manner.

One class of theories situates the roots of emotion and affect in bodily homeostasis (Damasio & Damasio, 2024; Man & Damasio, 2019), and/or interoceptive allostasis thereupon (Barrett, 2017), suggesting this route to quasi-affective artificial systems (Man & Damasio, 2019). On this view, valenced sensations of bodily warmth and cold are grounded in thermo-regulation (Barbosa Escobar et al., 2021). Postural gravity resistance is similarly valenced and relevant to emotion (Fuchs & Schlimme, 2009; Moradi & Mashak, 2013; Stocker et al., 2019). Postural slump is characteristic of depression (Feldman et al., 2020). The association of sensations of lightness with elation and joy, and of heaviness with sadness and depression are both familiar to experience and empirically observed (Hartmann et al., 2023). If balance is ancillary sensorimotor control, this valenced aspect is quite puzzling, but on a vital self-regulatory paradigm it would make more sense.

Whilst we typically think of balance as a problem for terrestrial animals such as humans, other animals face the same challenge of living with gravity in different contexts. Animal life began in the sea, a fluid medium dense enough to support a dense body (Nielsen, 1998). Many fish maintain neutral buoyancy by inflating a swim bladder to regulate their volume (Evans & Damant, 1928; Macaulay et al., 2020; Oppedal et al., 2020; Saunders, 1965), as in effect do scuba divers (Valenko et al., 2016). Failure to do so is a threat to viability and thriving (Oppedal et al., 2020). Other aquatic animals have their own solutions (Minamikawa et al., 1997; Newton & Potts, 1993). Terrestrial reproduction in air selected specific innovations of the amniote clade such as hard eggs and uterine gestation (Laurin & Reisz, 1995). Is there a general task definition for balancing which applies to both fish and human? Which encompasses gravity-resistance across examples and levels of description, as thermo-regulation encompasses adaptation from neurons to organism-level behaviour? A more general task is indeed apparent (our bold);

*"…the fetus is living in an aquatic environment, where the putative load receptors are not monitoring any weight.* ***Adaptation to weight*** *is therefore the main task confronting the sensorimotor system after birth."* (Massion, 1998) pp.469

Adaptation to weight is a more general and fundamental task than CoM/BoS stability or uprightness, and extends to multiscale mechano-regulation in general (Porazinski et al., 2015). Yet it has received relatively little focused attention as a task definition, presumably by dint of being broadly accepted as self-evident, and of being too vague to be of much practical use. This is unfortunate, because it is not quite correct and correcting it makes it useful. Conclusively, the weight of the foetus does not change much at birth, and so adaptation to weight is not a novel task at birth. Nonetheless, the *"putative load receptors"* will indeed register more load in the postnatal environment. This is because load is a function of *buoyancy,* and buoyancy does change over the perinatal transition, though not only at birth. Suspended in amniotic fluid there was no task, or the task was solved, but postnatally suspended in air, there is a task. This locates developmental task definition, as shown in the following sections. The fundamental task of terrestrial balance is *adaptation to negative buoyancy*, or more generally adaptation to non-neutral buoyancy. The next section outlines the key concept of buoyancy and formally defines the particular usage here.

## 3. Buoyancy matters

We are accustomed to thinking that things fall down because of gravity. This is true, yet an apple translates downwards in air, but upwards in water. The Earth's gravity, and the apple's weight, is constant between these cases. What changes is the density of the containing medium, and hence the apple's *buoyancy* (Turner, 1979). Buoyancy is a relational property describing the force imbalance between the gravitational acceleration of a massive body and of its fluid medium, or equivalently between gravitational acceleration and buoyant force, manifest as 'load'. If an object is denser than its fluid medium (e.g. an apple in air), it falls down; this is 'negative buoyancy'. The opposite is termed 'positive buoyancy' (e.g. an apple in water or a helium balloon in air). When density of object and containing medium is identical ('neutral buoyancy'), the object *maintains position relative to its medium* (other things remaining equal), because gravity is accelerating body and medium equally. Whilst gravity defines the vertical and provides the force, buoyancy defines whether a given massive body goes up or down.

Consider an inhomogeneous fluid planet close to equilibrium, relaxing under its own gravity. Matter will stratify according to density (Gvirtzman et al., 2016; Watts, 2001). If masses are not in their equilibrium density stratification, they will fall or rise until they equilibrate, as in oil separating from water; density gradients out of equilibrium will dissipate. Thus one fundamental biophysical challenge to mechano-regulation of the inhomogeneous body of an organism is to conserve a far-from-equilibrium collection of density gradients. 'Balancing' is the whole-body manifestation of this task.

The standard formal definition of buoyancy is $\rho_{Body}/\rho_{Medium}$, where a result >1.0 indicates negative buoyancy, 1.0 neutral buoyancy, and <1.0 positive buoyancy. Equation 1 reframes this definition to make it more convenient for our purposes and defines buoyancy ($B_{Body}$) as used throughout this paper. This is a more robust formulation as it avoids a divide-by-zero error when the medium is a vacuum; as such this formulation may be generally preferable.

$$1.\quad B_{Body} = 1 - \frac{\rho_{Medium}}{\rho_{Body}}$$

Reversing the denominator and taking the inverse situates the origin of $B_{Body}$ at neutral buoyancy. This is useful because $B_{Body}$ specifies buoyancy as a multiplicative gain term on gravitational acceleration. If $B_{Body}$ is 0.2 gravity will accelerate the body downwards relative to the medium, with the force of 20% of its mass (because 80% of the mass is 'insulated' from gravity by the containing medium). If $B_{Body}$ is -10, then gravity will accelerate the medium downwards relative to the body, and so the body upwards, with the force of 1000% of its mass. Note the signs are reversed; negative $B_{Body}$ corresponds to positive buoyancy. This mathematical reframing reflects the inherent biophysical cost of *distinguishing oneself* from ones medium (Ramstead et al., 2018; Schrödinger, 1992), here in terms of density. Section 5 unpacks this task definition in more detail and specifies the argument in an unambiguous formal manner. First, the next section establishes plausibility by reviewing evidence regarding amniotic regulation of foetal buoyancy during human pregnancy and its role in enabling proper morphogenesis and behavioural development.

## 4. Prenatal buoyancy in amniotic fluid

Intuitively, approximately equidense amniotic fluid supports the auto-construction of the growing embryo and foetus against gravity. More formally, equidense fluid suspension provides a three dimensional (3D) supporting volume in which all configurations of the contained body are unstable equilibria (Turner, 1979). This *poised* state prevents falling, and facilitates movement; exactly what is required of balancing. Instead, for a body with non-zero buoyancy ($B_{Body}$ in Eq. 1) equilibria are confined to 2D surfaces of the medium (the floor or surface), so external physical forces will work to squash a 3D body towards flatness and ultimately death. Accordingly, amongst the earliest post-implantation acts of the developing blastocyte is to form a fluid filled amniotic sac which envelopes the embryo/foetus throughout term (Cunningham et al., 2010). Within, the foetus is significantly insulated against gravitational acceleration because its mean density approximately matches that of its enveloping amniotic fluid (Boyd, 1933; Meban, 1983).

However, buoyancy is not perfectly neutral nor constant across term. Early (sub-1kg) human foetuses have a mean density around 0.971g/cm$^3$, increasing gradually to reach 1.021 g/cm$^3$ in the 3kg+ foetus (Meban, 1983). Amniotic fluid density also changes over term, but is slightly higher than that of water (1g/cm$^3$) (Underwood et al., 2005). Thus foetal buoyancy over term is determined by a complex, individual interplay of relative foetal-amniotic density. Amniotic fluid volume is actively regulated (Beall et al., 2007), but we have been unable to find evidence for or against regulation of amniotic fluid density. Measures reported by Meban suggest slightly positive buoyancy through early-mid term (Meban, 1983). However, observational estimates from ultrasound scans suggest neutral or slightly negative buoyancy during the second trimester (Hayat & Rutherford, 2018; Kurjak et al., 2005; Sekulić et al., 2005). As such, the trajectory of buoyancy over term remains underdetermined empirically, but may be important to many aspects of development and would benefit from more focused research.

Relative amniotic fluid volume decreases between weeks 20-26, and the growing foetus increasingly contacts the uterine walls, such that most foetuses are bearing 60-80% of their full body weight by week 26 according to (Sekulić et al., 2005). Polyhydramnios (too much amniotic fluid) at this stage can cause smaller, weaker foetal/neonatal bones (Ipek et al., 2016), which may reflect reduction in load-stimulated bone growth (Miller, 2025; Miller, 2003). During the third trimester the growing foetus is increasingly ‘swaddled’ by the limits of the uterine cavity. This period coincides with reduced motility, increased sleeping, and establishment of cortico-spinal projections (Einspieler et al., 2008; Eyre et al., 2000). The final transition is birth and a life compensating a steep body-medium density gradient.

The core organisation of the foetus is defined during the early term period of nearly-neutral buoyancy (Einspieler et al., 2008; Meban, 1983; Vinay et al., 2005). As a result, the foetus will develop extensive bodily and behavioural dependencies upon the buoyant force which holds it up; without buoyant support foetal posture would be crushed out of shape, as indeed can occur in oligohydramnios (Christianson et al., 1999). Similarly, early developing oscillatory movements embedded in the spinal pattern generators (Einspieler et al., 2008; Nishimaru & Kudo, 2000) which drive locomotor cycles under biomechanical and environmental constraints (Ijspeert & Daley, 2023) will implicitly ‘expect’ buoyant support, and are highly conserved across the lifetime (Dewolf et al., 2020; Rossignol et al., 2009). These are just a few examples, but the development of the whole body will be affected by buoyancy conditions.

One study has manipulated prenatal buoyancy directly, which manipulation altered the motor-developmental pathway in chick embryos (Bradley, 1997). In a computer simulated human foetus model, a comparator extrauterine simulation with buoyant support removed (foetus laying on a 'bed' surface) resulted in profoundly different sensorimotor organisation and behaviour (Yamada et al., 2013, 2016). The authors report disruption of modular organisation of the haptic and proprioceptive body map. Local cortical propagation of somatosensory input was curtailed in the extrauterine condition relative to intrauterine, and fewer cortical regions responded to specific body parts. Being loaded against the bed induced many physical couplings which caused correlations between body parts "increased correlations within body parts in tactile sensation and the decreased correlations across body parts in proprioception that are evoked by intrauterine embodied interactions facilitate the learning of body representations for specific single body parts". Resultant disruption of neural body-mapping hampered visual-somatosensory integration in a postnatal extension of the simulation. This in-silico study demonstrates the importance of the buoyant support provided by the uterine environment to proper whole-body scale calibration body-mapping and self-integration.

During the second and third trimester, the foetus is an active learner (Fagard et al., 2018; Yigiter & Kavak, 2006), learning smells, tastes and sounds (Kumar et al., 2023). It may develop sensorimotor predictions (Fagard et al., 2018; Hayat & Rutherford, 2018; Myowa-Yamakoshi, 2006; Zoia et al., 2007), for example of the contingencies of hand-face self-touch (Myowa-Yamakoshi, 2006; Reissland, 2013), whose accuracy will (implicitly) depend upon buoyant support of the arm's weight. Therefore, perinatal loss of buoyant support will cause distributed stressing and destabilisation of the bodily-behavioural milieu; a systemic *surprise* in the Bayesian sense (Friston, 2013; Ramstead et al., 2018). This recruits a multimodal assembly of subcomponents whose shared goal is to restore stability of the internal milieu (Bechtel & Bich, 2024a, 2024b; Bernard, 1957; Bettinger & Friston, 2023; Cannon, 1929). This shared goal entrains diverse subcomponents into behavioural coherence, cohering the diffuse physiological basis of the 'sense of balance' (Ferrè et al., 2013; Ferrè & Haggard, 2020; Nakamura et al., 2020) and its motor-postural counterpart (Ivanenko & Gurfinkel, 2018; Massion, 1998; Massion et al., 2004); resulting in a meaningful and distinct task-based sensorimotor modality of balancing.

For example, the human vestibular organ is anatomically mature from around 20 weeks post-conception (Jeffery & Spoor, 2004), and is proto-functional in utero (Ronca et al., 2008). However, when the foetus is neutrally buoyant vestibular activity will not be associated with falling. Perinatal loss of neutral buoyancy will introduce that association, exemplifying how that perturbation *recruits* subcomponents into an integrated sense of balance. Postnatally, diverse effects of load across the body will correlate with vestibular activity, which will therefore provide a key integrating sensory reference for both action and perception (Bliss et al., 2023; Gallagher et al., 2021; Lacquaniti et al., 2014). Indeed, mice engineered to have non-functional vestibular organs develop extensive disorders of self-integration (Le Gall et al., 2019). Other graviception and load sensing mechanisms (Duysens et al., 2000; Jansson et al., 2018; Kuldavletova et al., 2020; Ohlsson et al., 2020) will also be recruited and inform active balancing behaviour.

## 5. ‘Active buoyancy’ mechano-regulation

This section specifies the argument conceptually, and in simple formal terms to render the concepts unambiguously. The formal definition is purely descriptive/explanatory. The proposal is not that it is somehow internally represented in animals or should be in robots. Rather, the argument explains why and how animals are calibrated to apprehend and enact this task for their specific embodiment by prenatal development in approximately equidense fluid. Robots could be calibrated by the same developmental process, given a self-regulatory adaptive organisation with certain constraints and control capacities. The ‘value function’ for active buoyancy is not a formal term, but a function of the phenotype or robot; the difference between ‘being me’ in neutral buoyancy versus in non-neutral buoyancy, whatever ‘me’ may be.

Buoyancy (i.e. $B_{Body}$ in Equation 1) describes the mean density gradient at a massive body’s surface, and so the extent to which the organism ‘sticks out’ into the gravitational gradient. Maintaining a body-medium density gradient is costly in the same sense as maintaining a temperature gradient; the biophysical costs of living are precisely those of distinguishing oneself from one’s environment (Cannon, 1929; Ramstead et al., 2018; Schrödinger, 1992; Varela et al., 1974). Any such exposure (i.e. $B_{Body} \neq 0.0$) implies a proportional force debt which *will* extract energetic compensation.

Persistence in form and position requires compensating any body-medium density gradient, which means intervening on Equation 1. This can be done by changing $\rho_{Medium}$ (e.g. get in a pool), and/or changing $\rho_{Body}$. The latter means an *actuated* change $\rho_{Act}$ to body density. We denote this active, non-equilibrium approximation to buoyancy $B'_{Body}$ (Equation 2). $B'_{Body}$ will tend to zero no-matter-what, but the equilibrium trajectory leads to dissipation and death. The biophysical task of balance is actuating $\rho_{Act}$ such that $B'_{Body}$ tends to zero *whilst conserving* $\rho_{Body}$ within viability range. This demands a *non-equilibrium* trajectory compatible with the body’s persistence and incorporating the active component of work $\rho_{Act}$.

$$2.\quad B'_{Body} = 1 - \frac{\rho_{Medium}}{\rho_{Body+}\rho_{Act}}$$

The subset of viable trajectories $\rho^*_{Act}$ of $\rho_{Act}$ where $B'_{Body} \rightarrow 0.0$ whilst conserving $\rho_{Body}$ in viable range can be achieved in two ways; (1) by changing body volume ($V_{Act}$), as in a fish inflating a swim bladder, and/or (2) by changing body mass ($M_{Act}$). Body mass cannot easily be changed, but mass can ‘in effect’ be cancelled/grounded by levering a mechanical force field from the substrate surface normal across the body surface. Thus the task breaks down:

$$2.1\quad B'_{Body} = 1 - \frac{\rho_{Medium}}{(M_{Body} + M_{Act})/(V_{Body} + V_{Act})} \rightarrow 0.0$$

For animals living at the land-air interface, $M_{Act}$ is the main mode of work. Assuming constant volume (approximating a human in air), $M_{Act}$ must be sufficient, given local gravity, to cancel sufficient excess mass (i.e. $M_{Act} = -M_{Excess}$) such that the ‘effective’ density gradient $B'_{Body}$ tends to zero whilst conserving $\rho_{Body}$ in viable range;

$$3.\quad M_{Excess} = \left(1 - \frac{\rho_{Medium}}{\rho_{Body}}\right) M_{Body}$$

Equation 3 clarifies how buoyancy acts as multiplicative gain control on gravitational acceleration of body mass. This formulation just states that postural balance must compensate load $M_{Excess}$ (Duysens et al., 2000), taking into account that load is a function of buoyancy; load is in effect a physically given ‘error signal’ indicating deviation from neutral buoyancy. Achieving $\rho^*_{Act}$ by actively compensating mass as above is no longer neutral buoyancy *per se*, but effortful mechano-regulation which actively emulates neutral buoyancy by internally replacing ‘missing’ ambient mechanical (buoyant) force, thereby cancelling force from gravitational acceleration (Masani et al., 2013).

‘Missing’ has a specific meaning here. For any given configuration of a massive body, there is a mechanical force field at its surface which would be provided by the buoyant support of an equidense fluid medium, and which precisely supports body configuration under local gravity (Turner, 1979). ‘Missing’ buoyant force is defined by the difference between this notional neutral buoyancy force field and the actual current ambient buoyant force field. To persist in current form and vertical position, *any massive body must internally compensate the difference*. A rock on a substrate can achieve this by passive rigidity, but for a non-rigid body it will require negentropic work. Maintaining this neutral-buoyancy-equivalent mechanical force field is the *goal* of whole-body active balancing; for arbitrary morphologies, postures, and trajectories. For now, we descriptively term the work (i.e. balancing, verb) of actuating this force field ‘active buoyancy’, and the goal force field of balance (noun) ‘active neutral buoyancy’. Importantly for soft robotics and biological cases, active buoyancy applies to arbitrary morphologies and configurations, and does not make assumptions of rigidity.

For an agent in neutral buoyancy suspension, there are no gravity-imposed attractors or pendular actions, just those resonant dynamics intrinsic to the biomechanics of the body. Thus the agent will grow into, embody and expect its *unloaded* phenotype. The 3D supporting volume (any configuration is an unstable equilibrium with respect to gravity) means the consequences of self-movement are preserved, not erased by external physical forces. Resonant rhythmic movements can therefore emerge more easily (than in negative buoyancy conditions) through motor exploration (Einspieler et al., 2008). These are learned, for example by embedding in neural pattern generators (Melnyk & Pitti, 2018; Nishimaru & Kudo, 2000). This learned behavioural repertoire is then at least partially *fixed as normative*; some parameters which were previously plastic during calibration become permanent set-points (which can be dynamical) to be maintained forthwith, as observed for example in conserved spinal pattern generators (Rossignol et al., 2009). Collectively, these set-points proxy the neutral buoyancy goal condition they emerged in, by generating error signals in response to the imposition of negative buoyancy/load. These error signals recruit control resources and normatively direct their learning to minimise the error signal. Over the perinatal transition buoyancy is reduced, incurring error signals and challenging/training the agent to add sensorimotor control components to compensate this perturbation.

Formally, one can define a set of possible postural configurations for a given morphology $P_{Possible}$, and a subset $P^*_{Possible}$ for which the agent can actuate active neutral buoyancy in the current environment (these are identical sets for a neutrally buoyant agent with space to move). To a first approximation, the sensorimotor control task of regaining postnatal balance is to recover the identity of these sets by internally compensating missing ambient buoyant force; to *re-balance* the dynamical behaving body in its new environment (Ashby, 1947; Cannon, 1929; Einspieler et al., 2008, 2021; Stanojevic et al., 2011). This will only be partially successful; some configurations and trajectories will prove impossible to balance.

A newborn animal will thus work to *regain its balance* (i.e. recover the identity of $P_{Possible}$ and $P^*_{Possible}$), in as volitional a manner as neonates can be said to do anything. This work of balancing is as simple, natural and vitally necessary, and also as complex and volitional, as a baby seeking warmth or milk. It is not subsidiary to volitional action, but is *in itself* a primary task which must be enacted *regardless* of ongoing volitional actions, much like thermo-regulation. Active buoyancy (non-exclusively) drives infants to assert posture and movement in the postnatal environment, despite the bumps, falls and pain this process entails (Adolph et al., 2012; Craighero, 2024). The neonate will attempt to enact familiar prenatal postures and motor patterns, mostly failing at first. Over time, an animal may learn that small deviations from active neutral buoyancy can facilitate movement (Le Mouel & Brette, 2017), and equally how movement can facilitate active buoyancy. For a given morphology, some configurations and trajectories will be more effective than others (Bongard, 2014). Such asymmetries will *give rise* to 'uprightness' (Horstmann & Dietz, 1990; Ivanenko & Gurfinkel, 2018; Mergner et al., 2003) as *functional not prescriptive*; 'upright' is just those postural alignments with the vertical which best afford $P^*_{Possible} \rightarrow P_{Possible}$, and these can change for example with perturbations to the body (Figure 1c). This is quite a radical departure from current conceptions of the physiological macro-organisation of balancing.

Similarly, locomotion is not (initially) a goal in itself, but emerges more like a settling point (Wirtshafter & Davis, 1977); the form of behaviour which best stabilises the prenatally embedded oscillatory regimes given the constraints of the postnatal environment. Certain embodiments will be better for locomotion than others. In robotics these would need to be identified separately by hand or by evolutionary search in morphology space (Floreano & Keller, 2010; Harvey, 1997); active buoyancy addresses only balancing for a given agent.

The active neutral buoyancy force field must be continuously (approximately) maintained to achieve bodily-behavioural viability throughout life. As an animal develops postnatally, current capacity to maintain active neutral buoyancy is informed by ongoing experience of doing so; this ongoing calibration task is an important topic for focused future work, especially given anatomical and behavioural growth. Non-gravity mechanical perturbations may both support and threaten the maintenance of active neutral buoyancy; the substrate is a key enabler, whilst a shove may be a threat. Thus developing an interactive mastery of the mechanical support/threat afforded by surrounding mass becomes a key learning objective for skilled balance. For example, animals discover that CoM/BoS (Pollock et al., 2000; Van Dieën et al., 2024) is necessary for levering sufficient mechanical force from the substrate surface normal to enact active buoyancy mechano-regulation.

## 6. Discussion and conclusions

The present argument proposes to replace the current axiomatic definition of animate balance (static mechanical equilibrium) with a more general one (active neutral buoyancy). The latter accommodates movement, and subsumes the former as a special case (i.e. active neutral buoyancy with other things remaining equal). In proposing such a fundamental change, and in entailing a paradigmatic shift in how we conceive the physiological macro-organisation of balancing, this approach is conceptually radical. However, it is mostly compatible with existing work on the operational details of balancing, and is founded on inevitable biophysical requirements and observable developmental process. The key operational difference for robotics is that a period of initial development in equidense fluid is essential. During this period, the animal or robot must calibrate bodily and behavioural set-points to proxy the extant condition of neutral buoyancy. This calibration must be set (i.e prevent re-calibration to negative buoyancy norms) before loss of neutral buoyancy, such that the agent is permanently calibrated to actuate active neutral buoyancy; a specific form of innateness. This developmental process will address foundational task definition for balancing, for arbitrary morphologies and behaviours of arbitrary complexity.

Importantly, this generality means it can be used for soft and tensegrity robotics, fields where control paradigms which assume rigidity face challenges (Hawkes et al., 2021; Melnyk & Pitti, 2018; Rieffel & Mouret, 2018; Wen et al., 2020; Zhao et al., 2024). One answer to how a developmental system can become organised around an intrinsically unstable far-from-equilibrium goal (Heylighen, 2023) is 'co-homeostatic handover': that condition was previously stable under locally quasi-equilibrium conditions, but was then perturbed by a developmental transition. This challenges habituated norms and hence engenders compensatory adaptation which entrenches the status of the far-from-equilibrium goal as an attractor in the agent-world dynamical system (Di Paolo, 2010), which requires negentropic work to maintain.

There is more flexibility and richness to the 'active neutral buoyancy' goal than has been addressed in detail here. Thermo-regulation and body temperature are simple yet rich concepts subsuming much internal complexity and nested levels of function/description (Bechtel & Bich, 2024a; Benzinger, 1969; Cannon, 1929; Tan & Knight, 2018). Active buoyancy mechano-regulation and its goal of active neutral buoyancy are intended in the same sense. Small and temporary deviations are normal and necessary to movement, indeed skilled movement uses gravitational acceleration for effecting changes in position/configuration (Gillet et al., 2010; Ivanenko et al., 2007; Le Mouel & Brette, 2017). Modes of sensing and enacting the active neutral buoyancy force field may vary with body and control system design, just as 'maintain body temperature' has many potential solutions. The imperative to maintain active neutral buoyancy is non-exclusive as a driver of getting up and balancing. For example, infants may also observe adult standing and walking and attempt to copy it. However, it is proposed as the most fundamental and irreplaceable driver; animals do not need an external example to assert balance.

The main weakness of the current article is that it cannot cover all relevant aspects, most notably multiscale mechano-regulation of bodily density gradients, for example brain flotation (Theologou et al., 2022). Some important factors are beyond the scope of this article, but will be important to a more complete theory. The argument has focused on direct gravitational acceleration, and not addressed adaptation to wider gravitational effects such as atmospheric pressure (Pradillon & Gaill, 2006). Active buoyancy does not perfectly isolate gravity resistance because density and viscosity are associated. Adaptation to the change in viscosity from amniotic fluid to air will present a concurrent challenge to prenatally embedded sensorimotor habits, if maybe not a vital one (Underwood et al., 2005). This may be useful, in that the strength prenatally required to push through a viscous medium could be repurposed to resisting gravity. This may be an interesting and important interaction to model in future work.

The argument has set out the core conceptual and formal basis in an implementation-agnostic manner. Active buoyancy is a normative control policy (Körding, 2007) which may direct arbitrary optimisation algorithms; it specifies the quantity and target to optimise/satisfice for arbitrary bodies, *not* how it should be optimised by any given body. An important next step for future work is to develop concrete proof-of-concept robotic models which can demonstrate the theory, and act as models of biological cases to generate testable predictions. Active buoyancy can provide continuous error signals to guide action, via multiple channels and across multiple scales, minimisation of which will result in successful balance to the extent that bodily proxies are properly calibrated to the goal. However, for a complex robot/animal in a complex environment error minimisation will be non-trivial, with long-range temporal structure requiring more than momentary PID feedback control, and likely more than receding horizon control for reliable performance. Well-suited control and learning frameworks include homeostatic reinforcement learning (Hulme et al., 2019; Keramati & Gutkin, 2014; Man & Damasio, 2019; Yoshida, Daikoku, et al., 2024), self-regulatory active inference (Bettinger & Friston, 2023; Parr et al., 2025; Pezzulo et al., 2015; Tschantz et al., 2022), perceptual control theory (Cochrane & Nestor, 2025), ergodic control (Abraham et al., 2021; Fitzsimons & Murphey, 2022), and networks of adaptive frequency oscillators (Righetti et al., 2006; Righetti & Auke Jan Ijspeert, 2006). Having a general and encompassing task definition and associated error signal(s) means one can throw embodied intelligence, powerful machine learning algorithms, and arbitrary computational resources directly at the whole, isolated balancing task. This factor could enable a step-change in robotic balancing capability, including for soft robots. In turn, other behaviours could operate as if the body were unloaded, which is much easier.

In line with the notion that animal affect and valence are ultimately founded upon vital self-regulation (Damasio & Damasio, 2024), some have suggested that affective robotics may require similar organisation, and stressed the potential importance of robotic *softness* in this regard (Man & Damasio, 2019). However, robots do not have a life to lose (excepting perhaps bio-robots (Blackiston et al., 2023)), nor are they often responsible for maintaining their own body in more than a trivial sense, rendering the notion of vital self-regulation for robots somewhat moot (Di Paolo, 2010). Nonetheless, quasi-self-regulatory organisation can be modelled in artificial agents and robots (Ashby, 1960; Di Paolo, 2010; Pezzulo et al., 2015; Tschantz et al., 2022; Yoshida, Arikawa, et al., 2024).

Di Paolo's 'organism-inspired' robotics suggests that entrenched behavioural habits and their active conservation against perturbations could provide a behavioural level of self-regulatory organisation accessible to artificial agents. Such perturbations could be typical to development (as in the perinatal buoyancy transition), as well as atypical (Figure 1c). The stress they impose could form the "cognitive glue" binding disparate bodily mechanisms into coordinated activity to compensate the perturbation (Shreesha & Levin, 2024). This offers a longer-term adaptive medium in which to embed agent-level goal-proxies which are not merely determined by the form and short timescales of the sensorimotor loop, and hence are resilient to perturbations of it (Di Paolo, 2010). Di Paolo's in-silico experiments show how Braitenburg vehicle style agents (Braitenberg, 1986) can recover a habitual behaviour (positive phototaxis) despite a radical perturbation; left-right swapping of the visual sensors and associated wiring (which induces negative phototaxis). This approach offers a (quasi) self-regulatory framework for developmentally embedding proxies tracking the active neutral buoyancy goal as deeply into a robot's adaptive organisation as the robot's ability to sense and act upon its own body is deeply embodied. Here, balance would be both *cause and consequence* of behaviour and regulatory activity, closing the control loop. This may offer potential for more resilient, adaptive balance closer to that observed in animals, but dedicated research will be required to establish and exploit this potential.

The active buoyancy model is plausible given current knowledge, not least the evolutionary conservation of dense fluid suspension during prenatal development (Laurin & Reisz, 1995). To our knowledge, no extant species or lineage has evolved away from this reproductive adaptation. A growing body of research points to perinatal continuity of neuro-behavioural function between pre and postnatal life, and the conservation of core spinal motor patterns (Dewolf et al., 2020; Einspieler et al., 2008; Fagard et al., 2018; Kanazawa et al., 2023; Lacquaniti et al., 2014; Stanojevic et al., 2011; Sylos-Labini et al., 2020; Vinay et al., 2005). Precocial walkers such as piglets add motor components to rapidly adapt prenatally developed cyclic motor patterns to the postnatal environment and thereby conserve their stability (Vanden Hole et al., 2017). In altricial animals (such as humans) this adaptation is slower, but still relies on the same prenatally developed and postnatally conserved spinal and embodied components (Thelen et al., 1981; Thelen & Fisher, 1982), even across gaits (Dewolf et al., 2020).

(Dietz et al., 1989) reported that postural balance reactions at the ankle (as measured by electromyography) scaled linearly with buoyancy and were absent in neutral buoyancy, regardless of uprightness. Postural reactions in the ankle do not just respond to ankle loading, but enact the torque required to counteract gravitational acceleration (Masani et al., 2013), or equivalently to maintain active neutral buoyancy. Physiological adaptation also follows level of buoyant support over longer timescales; extended buoyant suspension induces losses in musculoskeletal condition similar to extended orbital microgravity (Tomilovskaya et al., 2019). Postural vertical is tilted forward from the gravitational vertical underwater, whilst habitual movement may nonetheless maintain CoM/BoS relations (Massion et al., 1995).

Thelen and colleagues hypothesised that biomechanical constraints prevent young human infants performing previously achievable stepping movement patterns. They found that those movement patterns can be elicited by providing ambient buoyant support artificially (Thelen et al., 1981; Thelen & Fisher, 1982). This finding perfectly demonstrates the *inter-changeability* of ambient buoyant force and the endogenously actuated force aspect of active buoyancy. This inter-changeability finds clinical application in aquatic therapy, which uses fluid buoyancy to insulate patients with sensorimotor damage or disorders from gravity loading (Becker, 2009; Gao et al., 2024). Aquatic therapy has also been used to support development in preterm neonates with some success (Aranha et al., 2022), and might also be applicable to gait training in other cases (Lorentzen et al., 2020).

The survival of preterm infants may appear to present a challenge to the active buoyancy theory. However, typically developing human foetuses lose neutral buoyancy around natal week ~20-26 (with individual variation), after which a foetus is bearing upwards of 60% of its body weight according to estimates from existing data given by (Sekulić et al., 2005). For a foetus to survive this (typical) loss of neutral buoyancy, it must be 'ready for heaviness' to a certain extent. The earliest age at which preterm infants survive is around 23 weeks (Myrhaug et al., 2019; Platt, 2014), so preterm survival limits coincide with the apparent loss of neutral buoyancy in healthy pregnancy, though more precise measurement of late-term foetal buoyancy are desirable. The biomechanical model of the 'preterm bone disease' places causality in insufficient load-stimulated bone growth (Miller, 2003). There is some evidence that polyhydramnios (too much amniotic fluid in late term) can cause reduced bone size and strength in full term newborns (Ipek et al., 2016). However, nutritional/mineral deficits are also relevant (Chinoy et al., 2019), and recently a multifactorial hypothesis has been proposed (Miller, 2025). Preterm infants with atypical general movements exhibit long term antigravity movement and motor deficits, whilst preterm infants with normal general movements do not (Miyagishima et al., 2016, 2018).

Though the active neutral buoyancy goal/task is ubiquitous, clearly not all organisms use the reproductive adaptation of formative neutral buoyancy suspension to apprehend it. For example, trees do not. Perhaps relatedly, tree trunks are characterised by rigidity/quasi-rigidity, and by being rooted into the substrate; notably different solutions from those of motile, non-rigid animals. Neutral buoyancy suspension may be practically necessary to the formative development of an animate, non-rigid behaving phenotype. Firstly, to extract the offspring phenotype from the substrate (so that it may move across the substrate). Secondly, for bootstrapping the growth of a non-rigid behaving body capable of autonomous motility and calibrating it to enable that motility by maintaining active neutral buoyancy across the lifespan.

The objective here is not a full and complete theory, but to sketch the beginnings of a vital mechano-regulation model of balance and stimulate a conversation which will lead to further theoretical development, empirical tests and practical applications. The argument has covered multiple domains and summarised some large bodies of literature very briefly, and cannot have done justice to the full range of extant opinion. Responses are invited to better unpack existing points of view, and how they may or may not relate to and align with these proposals. Many testable implications of the 'active buoyancy' model have been confirmed already (Bradley, 1997; Christianson et al., 1999; Dietz et al., 1989; Ipek et al., 2016; Masani et al., 2013; Thelen & Fisher, 1982; Yamada et al., 2013, 2016), and some further predictions are noted below. The present Perspective is intended as a theoretical foundation to establish the research programme suggested by these predictions.

Predictions

1. Foetuses across species will be approximately equidense with their individual amniotic fluid.
2. Large, extended perturbations to foetal buoyancy will be fatal and smaller/shorter ones may impose developmental pathologies. Typical individual trajectories of foetal buoyancy might be related to individual differences in postural development and expression.
3. Preterm infant survival and thriving will increase with increased individual experience of prenatal negative buoyancy and loading conditions.
4. Biomechanically realistic robots/simulations running the active buoyancy control policy will better model the directed efforts of neonatal ungulates to learn to get up, balance and walk than alternative control policies based on uprightness or secondary task rewards. Specifically, this approach will make the difference between random flailing during initial from-scratch learning, versus directed getting-up and staying-up.
5. The perinatal buoyancy transition will recruit physiological and behavioural subcomponents into a task-based modality manifest at the individual level in physiological, behavioural and neural measures. Such activity will emerge with perinatal loss of buoyancy, and tend to zero as buoyancy tends towards neutral.
6. Postural balance in positive buoyancy conditions (e.g. strong brine) with a ceiling as a substrate-replacement will mirror postural balance in negative buoyancy, probably with an interesting adaptation period reflecting reversed verticality. This might provide a useful paradigm for testing aspects of prediction 5, and would provide a direct test of the active buoyancy versus 'uprightness' hypotheses.

Acknowledgements

Many thanks to Luc Berthouze, Gustaf Gredebäck, Anna Ciaunica, Takashi Ikegami, Blake Richards, Karl Friston and the Theoretical Neurobiology Group for useful discussions and feedback.

Funding: None
Conflict of interest: None